%Paper: hep-th/9310032
%From: HAGENSEN%EBUBECM1.BITNET@FRMOP11.CNUSC.FR
%Date: Wed, 06 Oct 93 15:47:59 BCN

\newbox\leftpage \newdimen\fullhsize \newdimen\hstitle \newdimen\hsbody
\tolerance=1000\hfuzz=2pt
\def\bigans{b }
\def\answ{b }
%\message{ big or little (b/l)? }\read-1 to\answ
%
\ifx\answ\bigans\message{(This will come out unreduced.}
%AQUI\magnification=1200\baselineskip=16pt plus 2pt minus 1pt
\hsbody=\hsize \hstitle=\hsize %take default values for unreduced format
\else\def\apans{l }\message{ lyman or hepl (l/h) (lowercase]) ? }
\read-1 to \apansw\message{(This will be reduced.}
\let\lr=L
%AQUI\magnification=1000\baselineskip=16pt plus 2pt minus 1pt
\voffset=-.31truein\vsize=7truein
\hstitle=8truein\hsbody=4.75truein\fullhsize=10truein\hsize=\hsbody
%       Send local landscape command to laserprinter
\ifx\apansw\apans\special{ps: landscape}\hoffset=-.59truein% apple lw
  \else\hoffset=.05truein\fi% qms laserprinter
\output={\ifnum\pageno=0 %%% This is the HUTP version
  \shipout\vbox{\hbox to \fullhsize{\hfill\pagebody\hfill}}\advancepageno
  \else
  \almostshipout{\leftline{\vbox{\pagebody\makefootline}}}\advancepageno
  \fi}
\def\almostshipout#1{\if L\lr \count1=1
      \global\setbox\leftpage=#1 \global\let\lr=R
  \else \count1=2
    \shipout\vbox{\ifx\apansw\apans\special{ps: landscape}\fi %satisfies dvips
      \hbox to\fullhsize{\box\leftpage\hfil#1}}  \global\let\lr=L\fi}
\fi
%
%---------------------------------------------------------------------
\catcode`\@=11 % This allows us to modify PLAIN macros.
\newcount\yearltd\yearltd=\year\advance\yearltd by -1900

%
% restores pagenumbers
%
\def\draftmode{\def\draftdate{{\rm preliminary draft:
\number\month/\number\day/\number\yearltd\ \ \hourmin}}%
\headline={\hfil\draftdate}\writelabels\baselineskip=20pt plus 2pt minus 2pt
{\count255=\time\divide\count255 by 60 \xdef\hourmin{\number\count255}
        \multiply\count255 by-60\advance\count255 by\time
   \xdef\hourmin{\hourmin:\ifnum\count255<10 0\fi\the\count255}}}
%use this one instead of \Date on the preliminary draft
%it puts the current date on each page in big mode

%use \nolabels to get rid of eqn and ref labels in draft mode
\def\nolabels{\def\eqnlabel##1{}\def\eqlabel##1{}\def\reflabel##1{}}
\def\writelabels{\def\eqnlabel##1{%
{\escapechar=` \hfill\rlap{\hskip.09in\string##1}}}%
\def\eqlabel##1{{\escapechar=` \rlap{\hskip.09in\string##1}}}%
\def\reflabel##1{\noexpand\llap{\string\string\string##1\hskip.31in}}}
\nolabels
%
% tagged sec numbers
\global\newcount\secno \global\secno=0
\global\newcount\meqno \global\meqno=1
\def\newsec#1{\global\advance\secno by1\message{(\the\secno. #1)}
\xdef\secsym{\the\secno.}\global\meqno=1
%\ifx\answ\bigans \vfill\eject\else
\bigbreak\bigskip%\fi% (combination \goodbreak\bigskip\bigskip)
\noindent{\bf\the\secno. #1}\par\nobreak\medskip\nobreak}
\xdef\secsym{}
\def\appendix#1#2{\global\meqno=1\xdef\secsym{\hbox{#1.}}\bigbreak\bigskip
\noindent{\bf Appendix #1. #2}\par\nobreak\medskip\nobreak}
%
%       \eqn\label{a+b=c}       gives a displayed equation with number
%                               chosen consecutively within sections.
%     \eqnn and \eqna define labels in advance
%
\def\eqnn#1{\xdef #1{(\secsym\the\meqno)}%
\global\advance\meqno by1\eqnlabel#1}
\def\eqna#1{\xdef #1##1{\hbox{$(\secsym\the\meqno##1)$}}%
\global\advance\meqno by1\eqnlabel{#1$\{\}$}}
\def\eqn#1#2{\xdef #1{(\secsym\the\meqno)}\global\advance\meqno by1%
$$#2\eqno#1\eqlabel#1$$}
%
%                        footnotes
\newskip\footskip\footskip14pt plus 1pt minus 1pt %sets footnote baselineskip
\def\f@@t{\baselineskip\footskip\bgroup\aftergroup\@foot\let\next}
\setbox\strutbox=\hbox{\vrule height9.5pt depth4.5pt width0pt}
\global\newcount\ftno \global\ftno=0
\def\foot{\global\advance\ftno by1\footnote{$^{\the\ftno}$}}
%
%     \ref\label{text}
% generates a number, assigns it to \label, generates an entry.
% To list the refs on a separate page,  \listrefs
%
\global\newcount\refno \global\refno=1
\newwrite\rfile
\def\ref{\nref}
\def\nref#1{\xdef#1{[\the\refno]}\ifnum\refno=1\immediate
\openout\rfile=refs.tmp\fi\global\advance\refno by1\chardef\wfile=\rfile
\immediate\write\rfile{\noexpand\item{#1\ }\reflabel{#1}\pctsign}\findarg}
%       horrible hack to sidestep tex \write limitation
\def\findarg#1#{\begingroup\obeylines\newlinechar=`\^^M\pass@rg}
{\obeylines\gdef\pass@rg#1{\writ@line\relax #1^^M\hbox{}^^M}%
\gdef\writ@line#1^^M{\expandafter\toks0\expandafter{\striprel@x #1}%
\edef\next{\the\toks0}\ifx\next\em@rk\let\next=\endgroup\else\ifx\next\empty%
\else\immediate\write\wfile{\the\toks0}\fi\let\next=\writ@line\fi\next\relax}}
\def\striprel@x#1{} \def\em@rk{\hbox{}} {\catcode`\%=12\xdef\pctsign{%}}
\def\semi{;\hfil\break}
\def\addref#1{\immediate\write\rfile{\noexpand\item{}#1}} %now unnecessary
\def\listrefs{\immediate\closeout\rfile%\parindent=20pt
\baselineskip=14pt\centerline{{\bf References}}\bigskip{\frenchspacing%
%\catcode`\@=11
\escapechar=` \input refs.tmp }\nonfrenchspacing}
\def\startrefs#1{\immediate\openout\rfile=refs.tmp\refno=#1}
\def\figures{\centerline{{\bf Figure Captions}}\medskip\parindent=40pt}
\def\fig#1#2{\medskip\item{Fig.~#1:  }#2}
\catcode`\@=12 % at signs are no longer letters
%
%---------------------------------------------------------------------
%
\def\noblackbox{\overfullrule=0pt}
\hyphenation{anom-aly anom-alies coun-ter-term coun-ter-terms}
\def\inv{^{\raise.15ex\hbox{${\scriptscriptstyle -}$}\kern-.05em 1}}
\def\dup{^{\vphantom{1}}}
\def\Dsl{\,\raise.15ex\hbox{/}\mkern-13.5mu D} %this one can be subscripted
\def\dsl{\raise.15ex\hbox{/}\kern-.57em\partial}
\def\tr{{\rm tr}} \def\Tr{{\rm Tr}}
\def\lspace{\ifx\answ\bigans{}\else\qquad\fi}
\def\lbspace{\ifx\answ\bigans{}\else\hskip-.2in\fi} % $$\lbspace...$$
\def\boxeqn#1{\vcenter{\vbox{\hrule\hbox{\vrule\kern3pt\vbox{\kern3pt
        \hbox{${\displaystyle #1}$}\kern3pt}\kern3pt\vrule}\hrule}}}
\def\mbox#1#2{\vcenter{\hrule \hbox{\vrule height#2in
                \kern#1in \vrule} \hrule}}  %e.g. \mbox{.1}{.1}
\magnification 1200

\def\a{\alpha}  \def\b{\beta} \def\g{\gamma} \def\G{\Gamma}
\def\d{\delta} \def\D{\Delta} \def\ee{\epsilon} %\def\ee{\varepsilon}
\def\z{\zeta} \def\th{\theta} \def\TH{\Theta} \def\tth{\vartheta}
\def\k{\kappa} \def\l{\lambda} \def\L{\Lambda} \def\m{\mu} \def\n{\nu}
\def\cs{\xi} \def\Cs{\Xi} \def\p{\pi} \def\P{\Pi} \def\r{\rho}
\def\s{\sigma} \def\S{\Sigma} \def\t{\tau} \def\y{\upsilon}
\def\Y{\upsilon} \def\f{\phi} \def\F{\Phi} \def\x{\chi} \def\ps{\psi}
\def\Ps{\Psi} \def\o{\omega} \def\O{\Omega} \def\vf{\varphi}
\def\pa{\partial} \def\da{\dagger} \def\dda{\ddagger}

%REFERENCES

\ref\wilson{K.  Wilson and J. Kogut, {\it Phys.  Repts.} {\bf
12C}(1974)75\semi K. Wilson, {\it Rev.  Mod. Phys.} {\bf 47}(1975)773.}
\ref\wegner{F.J.  Wegner and A. Houghton, {\it Phys.  Rev.} {\bf
A8}(1972)401.}
\ref\polchinski{J. Polchinski, {\it Nucl. Phys.} {\bf B231}(1984)269.}
\ref\hasenfratz{A.  Hasenfratz and P. Hasenfratz, {\it Nucl.  Phys.}
{\bf B270}(1986)687.}
\ref\margaritis{A.  Margaritis, G. Odor and A. Patkos, {\it Z.
Phys.} {\bf C39}(1988)109.}
\ref\zamolodchikov{A.B. Zamolodchikov, {\it JETP Lett.} {\bf
43}(1986)730; {\it Sov. J. Nucl. Phys.} {\bf 46}(1987)1090.}
\ref\osborn{J.L. Cardy, {\it Phys. Lett.} {\bf B215}(1988)749\semi
H.  Osborn, {\it Phys.  Lett} {\bf B222}(1989)97\semi
I. Jack and H. Osborn, {\it Nucl.  Phys.} {\bf B343}(1990)647\semi
A. Cappelli, D. Friedan and J.I.  Latorre, {\it Nucl.  Phys.} {\bf
B352}(1991)616\semi A. Cappelli, J.I.  Latorre and X. Vilas\'\i
s-Cardona, {\it Nucl.  Phys.} {\bf B376}(1992)510.}
\ref\felder{G.  Felder, {\it Comm.  Math.  Phys.} {\bf 111}(1987)101.}
\ref\wallace{D.J. Wallace and R.K.P. Zhia, {\it Ann. Phys.} {\bf
92}(1975)142.}
\ref\cargese{C.  Itzykson and J.M.  Drouffe, {\it Statistical Field
Theory}, vol.1, Cambridge University Press (1989); cf. also several
references in {\it Phase Transitions} (Carg\`ese 1980), eds.  M. Levy,
J.C.  Le Guillou, J. Zinn-Justin.}

\def\footnoterule{\kern-3pt \hrule width \hsize \kern2.6pt}
\pageno=0
\footline={\ifnum\pageno>0 \hss\folio\hss \else\fi}

\centerline{\bf Gradient Flows From An Approximation To The Exact
Renormalization Group \footnote{*} {\sevenrm This work is supported in
part by funds provided by AEN 90-0033 Grant (Spain), and by M.E.C.
(Spain).}}
\vskip20pt

\centerline{ Peter~E.~Haagensen,
Yuri~Kubyshin\footnote{$^1$}{\sevenrm On leave of
absence from Nuclear Physics Institute, Moscow State University, 119899
Moscow, Russia.}, Jos\'e~I.~Latorre, Enrique~Moreno}
\vskip 20pt
\centerline{\sl Department d'Estructura i Constituents de la
Mat\`eria}
\centerline{\sl Facultat de F\'\i sica, Universitat de Barcelona}
\centerline{\sl Diagonal, 647~~08028 Barcelona, SPAIN}
%\vskip36pt

\vfill

\centerline{\bf ABSTRACT}
\midinsert
\baselineskip=13pt plus 1pt
\smallskip
Through appropriate projections of an exact renormalization group
equation, we study fixed points, critical exponents and nontrivial
renormalization group flows in scalar field theories in $2<d<4$.  The
standard upper critical dimensions $d_k={2k\over k-1}$, $k=2,3,4,\ldots$
appear naturally encoded in our formalism, and for dimensions smaller
but very close to $d_k$ our results match the $\ee$-expansion.  Within
the coupling constant subspace of mass and quartic couplings and for any
$d$, we find a gradient flow with two fixed points determined by a
positive-definite metric and a $c$-function which is monotonically
decreasing along the flow.
\endinsert
\vfill
\line{UB-ECM-PF\#93-20 \hfil October 1993}
\eject

\newsec{Introduction}

Wilson's exact renormalization group\wilson\ provides a functional
differential equation which dictates the way short-distance physics gets
integrated into a long-distance effective action.  This equation and its
analogues (Wegner-Houghton\wegner\ and Polchinski\polchinski ) are
certainly powerful but too complex to be of practical use.  Different
approximations and projections have been devised to bring these
equations to more workable, yet non-perturbative, settings.

In this Letter we elaborate on a projection of the Wegner-Houghton exact
renormalization group equation due to Hasenfratz and
Hasenfratz\hasenfratz.  The basic idea is to focus on the evolution of
the zero mode of the scalar field, with the highest momentum modes being
integrated out into a self-interaction term of the constant mode.  This
setting retains non-linearities from the original equation and is
amenable to both analitical and numerical studies.  Extending previous
work along the same lines\margaritis , we analyze the fixed points of
scalar theories for arbitrary dimensions.  We then address the
renormalization flows between fixed-points and, in particular, the
critical exponents.  This somewhat conventional piece of work is then
completed with a new study on the irreversibility of the flows.  It is
an open question whether renormalization group flows are gradient.  A
theorem due to Zamolodchikov\zamolodchikov\ ensures such a property in
two dimensions but only inconclusive work has been done in higher
dimensions (see, for instance, \osborn ). Here, with some guesswork and
brute force we are able to show that at first non-trivial order in our
approach the flow is indeed gradient and thus irreversible.  It is
determined by a positive-definite metric in coupling constant space and
a $c$-function which is monotonically decrasing along the flow, and it
connects a unique Gaussian fixed point to a unique Wilson fixed point
for any $2<d<4$.

\bigskip
\newsec{Projection of the Wegner-Houghton Equation}

The basic flow equation we will use is that of Wegner and
Houghton\wegner , while our notation follows that of \hasenfratz . The
basic and fairly intuitive procedure of Wegner and Houghton is to
consider a generic scalar field theory action $S$, with an UV momentum
cutoff $\L_0$.  Starting from that, one then performs in the path
integral an integration only over the outermost infinitesimal momentum
shell, with momenta $e^{-t}\L_0\le q\le \L_0$, and $t$ small.  An
effective action will result for the unintegrated fields, now with a
slightly smaller cutoff, $\L =e^{-t}\L_0$.  Momenta in this effective
action are then rescaled so their range again becomes $0\le q\le \L_0$,
and with that the fields themselves and the measure in the action will
scale with their appropriate scaling dimensions.  When all this is put
together, a differential equation results describing how the effective
action changes as this second cutoff is lowered and more and more
momentum degrees of freedom are integrated out.  We do not give the
derivation of this exact renormalization group equation, since it is
presented in detail in \wegner , but instead only state the final
result:
\eqn\erg{\eqalign{
{\pa S\over\pa t}&={1\over 2t}{\int_q}^{'}\left\{\ln
{\pa^2 S\over\pa\f (q)\pa\f (-q)}-{\pa S\over\pa\f (q)}
{\pa S\over\pa\f (-q)}\left({\pa^2 S\over\pa\f (q)\pa\f
(-q)}\right)^{-1}\right\}\cr
&-\int_q q_\m\f (q)\pa_{q_\m}^{'}{\pa S\over\pa\f (q)}+dS+
(1-{d\over 2}-\eta )\int_q \f (q){\pa S\over\pa\f (q)}+{\rm const.}\, ,}}
where the prime in the first integral above indicates integration only
over the infinitesimal shell of momenta $e^{-t}\L_0\le q\le \L_0$, and
the prime in the derivative indicates that it does not act on the
$\d$-functions in $\pa S/\pa\f (q)$.

By projecting onto the constant mode $\f (0)$ of $\f (q)$, the above
exact renormalization group equation is considerably simplified, and
becomes the flow equation only for the effective potential of the theory
(this, of course, also projects out some terms which could contribute to
the flow of the effective potential, but the fact that we are still able
to find a very rich structure in the ensuing flow motivates this
projection).  We furthermore use the approximation used in
Ref.\hasenfratz , constraining the effective action to have no other
derivative pieces than the canonical kinetic term, that is, in
coordinate space:
\eqn\approx{S=\int d^d\!x\,\left\{ {1\over 2}(\pa_\m \f )^2+V(\f )
\right\}\, .}
This leads to the following flow equation for the effective potential:
\eqn\vflow{\dot{V}(x,t)={A_d\over 2}\ln (1+V''(x,t))+d\cdot
V(x,t)+(1-{d\over 2}-\eta )xV'(x,t)+{\rm const.}\, ,}
where $A_d/2=[(4\p )^{d/2}\G (d/2)]^{-1}$, the dot is a scale derivative
$\pa /\pa t$, $x$ is the constant mode $\f (0)$, and we again refer the
reader to the derivation in Ref.\hasenfratz . In the approximation we
are using, Eq. \approx , we actually leave no room for a wavefunction
renormalization, and this turns out to imply that $\eta =0$ above.  For
greater ease of calculations, we will actually study the equation for
$f(x,t)=V'(x,t)$, trivially found from the above:
\eqn\fflow{\dot{f}(x,t)={A_d\over 2}{f''(x,t)\over
(1+f'(x,t))}+(1-{d\over 2})xf'(x,t)+(1+{d\over 2})f(x,t)\, ,}
with $\eta$ already set to 0. We remark here that the constant $A_d$ can
be absorbed by a rescaling of $x$, thus disappearing from the equation
above, a fact we will make use of later.  This is a reflection of
universality in Eq. \fflow, whereby the shape of $f^{*}$ will depend on
$A_d$ but the critical exponents will not.  This is the starting point
of our calculations.  From here one can proceed either by investigating
numerical solutions\hasenfratz\ or by analytical means.  We choose the
latter, where we will use the following polynomial approximation for
$f(x,t)$:
\eqn\fpoly{f(x,t)=\sum_{m=1}^M\, c_{2m-1}(t)x^{2m-1}\, ,} with $M$ an
arbitrary integer (and, naturally, better approximations will have
larger $M$), and where only odd powers are chosen because we want the
potential $V$ to be reflection-symmetric.  This approximation has been
widely used in the past (cf.\margaritis ,\felder ).

\newsec{Fixed Points}

With Eqs.\fflow\ and \fpoly\ as our starting point, our first objective
is to determine the allowed fixed point solutions and their properties.
This is easily done by substituting a polynomial fixed-point solution
\eqn\fspoly{f^{*}(x)=\sum_{m=1}^M\, c_{2m-1}^{*}x^{2m-1} }
with finite but arbitrarily large $M$ into the fixed-point equation,
i.e., Eq.\fflow\ with $\dot{f}=0$.  A Taylor expansion in $x$ then leads
to a set of $M$ nonlinear algebraic equations, of the form
\eqn\w{\eqalign{&w_1(c_1^{*},c_3^{*})=2c_1^{*}+{3A_dc_3^{*}\over
1+c_1^{*}}=0\cr &w_2(c_1^{*},c_3^{*},c_5^{*})=(4-d)c_3^{*}+
{10A_dc_5^{*}\over (1+c_1^{*})}- {9A_dc_3^{*}{}^2\over
(1+c_1^{*})^2}=0\cr &{}~~~~ \vdots \cr &w_{M-1}(c_1^{*},c_3^{*},\ldots
c_{2M-1}^{*})=0\cr &w_M(c_1^{*},c_3^{*},\ldots c_{2M-1}^{*})=0\, ,}}
which can {\it always} be solved exactly and recursively up to
$w_{M-1}$, giving $c^{*}_3, c^{*}_5,\ldots c^{*}_{2M-1}$ as a function
of $c^{*}_1$.  That is substituted in $w_M$, which then becomes a
polynomial of order $M$ in $c^{*}_1$ with the form:
\eqn\wn{w_M=k(d)\, c^{*}_1(\a_0(d)+\a_1(d)c^{*}_1+\ldots +\a_{M-1}(d)
c^{*}_1{}^{M-1})=0\, ,}
with $\a_i(d)$ being polynomials of order $M-1$ in $d$. (Note that
$c^{*}_1=0$ ( $\Rightarrow c^{*}_{i>1}=0$) is always a solution for any
$d$. This is the Gaussian fixed point.) As an example, for $M=6$, we
find:
\eqn\alphas{\eqalign{ k(d)&={1\over 155925A_d^5}\cr \a_0(d)&= (d -4)(d
-3)(3d -8)(2d-5)(5d -12 )\cr
\a_1(d)&=-699456+899960d-436386d^2+95973d^3-8651d^4+150d^5\cr
\a_2(d)&=-8763072+9018200d-3349144d^2+527126d^3-30234d^4+300d^5\cr
\a_3(d)&=-31764096+27691776d-8408924d^2+1030750d^3-43166d^4+300d^5\cr
\a_4(d)&=-42872880+32815292d-8479920d^2+851409d^3-28049d^4+150d^5\cr
\a_5(d)&=-19261320+13251980d-2990962d^2+254333d^3-6903d^4+30d^5\cr \,
.}}

For any given $M$, a complicated phase space of solutions $(d,c^{*}_1)$
can be found and plotted numerically.  We have done this up to $M=7$,
and the important aspects of these solutions can be summarized as
follows:

{\it i)} The first important feature is that, rather unexpectedly,
$\a_0(d)$ always factorizes into \eqn\fa{\prod_{m=2}^M\, (d-d_m)
=\prod_{m=2}^M \, (d-{2m\over m-1})=(d-4)(d-3) (d-{8\over3})\cdots\, .}
This means that at the upper critical dimensions $d=d_k, k=1,2,\ldots$,
$c_1^{*}=0$ is actually a double solution to $w_M=0$, which indicates a
branching of fixed-point solutions below these critical dimensions.
This is in perfect agreement with the multicritical fixed-point
solutions known to exist below these dimensions.

{\it ii)} To further corroborate the above, an $\ee$-expansion of
Eqs.\w\ and \wn\ about any critical dimension (with $M=k$) also leads to
the known $\ee$-expansion solution given by Hermite polynomials, that
is, for $d=d_k-\ee$, \eqn\hermi{f^{*}(x) =\k_k\ee H_{2k-1}(x/\l_k
)+{\cal O}(\ee^2) ,~~~~\l_k =\sqrt{{2{A_d}_k\over d_k-2}}\, ,} where
$\k_k$ is a constant depending on $d_k$ (for instance, for $k=2$,
$d_2=4$, and $\k_2=\sqrt{A_2}/72$).  Note that a simple $\ee$-expansion
of Eq.\fflow\ will lead to a linear equation and thus cannot furnish
this constant $\k_k$.  At higher orders in $\ee$ we expect our results
not to agree with the standard $\ee$-expansion since the present
approximation does not allow for wavefunction renormalization.

{\it iii)} When $M$ is increased by $1$, the first $M-1$ equations in
\w\ remain unchanged. The solution to the last one seems,  as far as we
have investigated, to lead to convergence of previous solutions as $M$
gets larger (see also \margaritis).

{\it iv)} The lower the dimension, the less trustworthy is this
approximation or, conversely, the larger is the $M$ needed.  Altogether,
we find that, for any $d$, some solutions represent true fixed points
while others are spurious.  For lower dimensions, the number of true
nontrivial fixed points increases, but so does the number of spurious
solutions.

\newsec{Flows and Critical Exponents}

Having found particular fixed point solutions in some approximation $M$,
we can now study how the renormalization flow approaches these solutions
by determining the critical exponents. To find them, we study small
$t$-dependent departures from some fixed-point profile $f^{*}(x)$:
\eqn\ffg{f(x,t)=f^{*}(x)+g(x,t)\, ,}
where again a polynomial {\it ansatz} is chosen for $g(x,t)$:
\eqn\gxt{
g(x,t)=\sum_{m=1}^M\, \d_{2m-1}(t)x^{2m-1}\, .}
When \ffg\ is substituted in Eq.\fflow\ and only linear terms in $g$ are
kept, the resulting equation is
\eqn\gdot{\dot{g}={A_d\over 2}\left[
{1\over (1+f^{*}{}')}g''-{f^{*}{}''\over (1+f^{*}{}')^2}g'\right]
+(1-{d\over 2})xg'+(1+{d\over 2})g\, .}
With the {\it ansatz} \gxt\ and matching powers of $x$ in \gdot , we
then find:
\eqn\ddot{\dot{\d}_i=\sum_{j=1}^M\, \O_{ij}(c^{*},d)\, \d_j\, ,}
where $\O_{ij}$ is an $M\times M$ matrix which depends on the input
values $c_i^{*}$ and $d$. For higher $M$, the entries are rather long
and unwieldy; here we present as an example $\O_{ij}$ for the $M=3$
case:
\eqn\omm{
\Omega_{i j}=\pmatrix{
  2-{6 {A_d} {c_3^*}\over (1+{c_1^*})^2}&
	     {6{A_d}\over 1+{c_1^*}}& 0\cr
  {4{A_d}(9{c_3^*}^2-5{c_5^*}(1+{c_1^*}))\over (1+{c_1^*})^3}&
	      (4-d)-{36 {A_d} {c_3^*}\over (1+{c_1^*})^2}&
	      {20 {A_d}\over 1+{c_1^*}}\cr
  {-18{A_d}\ {c_3^*}(9{c_3^*}^2-10 {c_5^*}(1+{c_1^*}))\over (1+{c_1^*})^4}&
       {18 {A_d}(9 {c_3^*}^2-5{c_5^*}(1+{c_1^*}))\over (1+{c_1^*})^3}&
	(6-2d)-{90 {A_d} {c_3^*}\over (1+{c_1^*})^2}\cr
		      } }

The critical exponents will be given by the characteristic frequencies
of the Eq.\ddot , i.e. the eigenvalues of $\O$.  For the fixed points
continuously connected to the only nontrivial fixed point above $d=3$,
we have calculated the critical exponents numerically up to $M=7$ for
dimensions between $2$ and $4$ in steps of $0.1$.  For $2.9\le d\le 4$
our results are plotted in Fig. 1. It is worthwhile noting that as
$d\rightarrow 4$, the critical exponents merge with the tower of
canonical dimensions $(2,0,-2,-4,\ldots )$, which are precisely the
critical exponents of the trivial Gaussian theory at $d=4$ (i.e., the
canonical dimensions of the ($\f^2,\f^4,\f^6,\ldots $) couplings in
$d=4$).  This is an indication in our setting of the existence of a
unique (Gaussian) fixed point at $d=4$.  We furthermore note that for
$d=3$ our two leading exponents \eqn\expnts{\eqalign{\n
={1\over\o_1}&=0.656\cr \o_2 &=-0.705}} match fairly well results gotten
by other methods (in \hasenfratz , $\n =0.687$ and $\o_2 =-0.595$; field
theory calculations, high temperature expansions and Monte Carlo
methods\cargese\ all yield $\n =0.630\pm 0.003$ and $-1.0<\o_2<-0.5 $).

It is also possible to perform an $\ee$-expansion on Eq.\gdot\ around a
critical dimension $d=d_k$.  We have done this for $d=d_2=4$,
calculating the critical exponents analitically in $\ee$ to ${\cal
O}(\ee^2 )$ and for all operators, and the procedure is identical for
all higher multicritical points.  To simplify our calculations we make
use of the universality of the critical exponents in $A_d$ and set it to
one.  Then, we substitute Eq.\hermi\ for $f^{*}$ in Eq.\gdot\ (and here
it is important to have the correct coefficient $\k_k$ in Eq.\hermi , as
well as the ${\cal O}(\ee^2 )$ terms), and make the following {\it
ansatz} for $g(x,t)$:
\eqn\eexp{g(x,t)= {\rm exp}\{ (\o_{\ell}^{(0)}
+\ee\o_{\ell}^{(1)}+\ee^2\o_{\ell}^{(2)})t\}\
(g_0(x)+\ee g_1(x)+\ee^2 g_2(x))\, ,}
where $\o_{\ell}^{(0)}=2(2-\ell ), \ell$ integer $\ge 1$, is any one of
the critical exponents at $d=4$.  At zeroth order in $\ee$, one finds
$g_0(x)\sim H_{2\ell-1}(x)$.  At first order in $\ee$, using the
previous solution for $g_0(x)$, one finds an equation of the form:
\eqn\omone{\eqalign{ {\cal H}_\ell g_1(x)&\equiv \left( {d^2\over
dx^2}-2 x {d\over dx}+2(2 \ell -1)\right) g_1(x) \cr &=\left(\o_{\ell}
^{(1)}+({1\over 2}+ {(2\ell -1)(2\ell -3)\over 6})\right) g_0(x)+\cdots
\, ,}}
where ${\cal H}_\ell$ is the Hermite operator which annihilates
$H_{2\ell-1}(x)$ and the dots are other Hermite polynomials.  Since
$g_0(x)$ on the r.h.s. is precisely this Hermite polynomial, its
coefficient must vanish for the equation to be consistent.  This
determines $\o_{\ell}^{(1)}$.  At next order, $\o_{\ell}^{(2)}$ is
determined in essentially the same way, that is, by canceling on the
r.h.s. a zero mode of the Hermite operator which appears on the l.h.s..
The final result is:
\eqn\fullom{ \o_\ell =2(2-\ell )-{\ee\over 2}\left( 1+{(2\ell -1)(2\ell
-3)\over 3} \right) +{2\ee^2\over 9}\ell(2\ell -1)(2\ell -3) +{\cal O}
(\ee^3)\, ,}
where $\ell =1,2,3,..$ labels the exponents for the different operators.
At order $\ee$ this agrees exactly with standard field theory
calculations in the $\ee$-expansion\cargese . At next order, as
announced above, our result differs slightly from the standard one due
to the absence of wave function renormalization and the truncation of
the exact renormalization group equation itself.  By a somewhat more
cumbersome calculation it is also possible to find the ${\cal O}(\ee )$
correction to the critical exponents analytically in $k$ for a generic
multicritical point.  The answer is:
\eqn\lamdagen{\o_{k,\ell}=2{(2-\ell)\over (k-1)}-\ee\left(\ell -1-2 (k-1
){(2 \ell)!\ k! \over (2\ell -k)!\ (2k)!}\right)+{\cal O}(\ee^2)\ . }

Critical exponents only characterize the flow very close to a particular
fixed point.  Another option we have is to study the flow globally by
substituting Eq.\fpoly\ directly into Eq.\fflow . Matching powers of $x$
in a Taylor expansion leads to coupled nonlinear flow equations for
$c_i(t)$ in the form:
\eqn\cdt{\dot{c}_i=w_i(c)\, ,~~~~i=1~~{\rm to}~~M\, ,}
where the $w_i(c)$ are given in Eq.\w . Arguably, a polynomial {\it
ansatz} does introduce a perturbative element into the essentially
nonperturbative nature of renormalization flows between distant fixed
points, and our approximation very likely misses some features of the
true flow.  However, we believe that, again, the sensible and rich
structure that emerges does justify the simplification.

We have solved the nonlinear flow \cdt\ numerically with $M=3$ in $d=3$:
\eqn\numflo{\eqalign{ \dot{c}_1&=2c_1+{3\over 2\p^2}{c_3\over 1+c_1}\cr
\dot{c}_3&=c_3-{9\over 2\p^2}{c_3^2\over (1+c_1)^2}+{5\over\p^2}
{c_5\over 1+c_1}\cr \dot{c}_5&={27\over 2\p^2}{c_3^3\over
(1+c_1)^3}-{45\over 2\p^2} {c_3c_5\over (1+c_1)^2}\, .}}
The $(c_1(t),c_3(t))$ subspace of that flow is shown in Fig. 2. We note
there the presence of a Gaussian (at $(0,0)$) and a Wilson fixed point,
and a unique trajectory leading from the former to the latter.  To
determine that this flow is gradient and permits a $c$-function
description is the object of the next section.

\newsec{c-Function}

We now study some features of the geometry of the space of local
interactions.  If the beta functions of a theory can be written as a
gradient in the space of coupling constants,
\eqn\grad{ \beta ^{i}(c)= - g^{i j}{\partial {\cal C}\over \partial
c_{j}} }
where $g^{i j}$ is a positive-definite metric, we know that the set of
renormalization flows becomes irreversible\wallace . In such a case,
there exists a function $\cal C$ of the couplings which is monotonically
decreasing along the flows:
\eqn\cmd{ {d {\cal C} \over {dt}}= \beta_{i} {\partial {\cal C} \over
\partial c^{i}}= - g^{i j} {\partial {\cal C} \over \partial c^{i}}
{\partial {\cal C} \over \partial c^{j}}\leq 0, }
making their irreversibility apparent, so that recurrent behaviors such
as limit cycles are forbidden.  In two dimensions it is possible to
prove that the fixed points of the flow are the critical points of $\cal
C$ and that the linearized RG generator in a neighborhood of a fixed
point is symmetric with real eigenvalues (the critical exponents).

The renormalization group flows found in the previous section are all
well-behaved.  Therefore it becomes natural to ask whether these flows
are gradient, {\it i.e.,} whether there exists a globally defined
Riemannian metric $g_{i j}$ and a non-singular potential ${\cal C}$
satisfying Eq. \grad.  The general solution for an arbitrary number of
couplings $M$ would be extremely difficult.  However, we find that it is
possible to treat the case $M=2$, namely, the subspace of mass and
quartic couplings.  The beta functions corresponding to the two
couplings $c_1,c_3$ are given in Eq. \numflo\ (where we restrict to
$c_5=0$).  Because of the positivity of $c_3$ ($c_3$ is the coefficient
of $\phi ^4$ in $V$ and is required to be positive for stability of the
path integral) it is appropriate to make the following coupling constant
reparametrization:
\eqn\changecor{ \eqalign{ c_1 & \to m^2=c_1 \cr c_3 & \to \lambda^2 = 6
A_d\ c_3\, . \cr } }

In these new variables the beta functions take the form
\eqn\nflows{ \eqalign{ {d\ m^2 \over dt} & = 2 m^2 + {1\over 2}{\lambda
^2 \over (1+m^2)} \cr {d\ \lambda \over dt} & = {(4-d)\over 2}\,
\lambda-{3\over 4}{\lambda ^3 \over (1+m^2)^2} \cr } }
and the fixed points become
\eqn\nfp{ \eqalign{ {\rm Gaussian:}\ \ \ \ \ \ & (m^2_{G},
\lambda_{G})=(0,0) \cr {\rm Wilson:}\ \ \ \ \ \ & (m^2_{W},
\lambda_{W})=(-{4-d\over 10-d},{\sqrt{24 (4-d)}\over 10-d} )\, . \cr } }
Note that the Wilson fixed point merges with the Gaussian one at $d=4$,
similarly to the situation in Sec. 4. Now, by trial and error and
considerable guesswork, the following solution to Eq. \grad\ can be
found:
\eqn\cfun{ {\cal C}(m^2,\lambda)={1\over 2} (1+m^2)^4 -{2\over
3}(1+m^2)^3 + {1\over 4} \lambda ^2 (1+m^2)^2 - {3\over 16} {\lambda ^4
\over (4-d)} } and \eqn\metrig{ g^{i j}={1\over (1+m^2)}\pmatrix{ 1& 0
\cr 0 & 4-d \cr}. }
${\cal C}(m^2,\lambda)$ has the expected properties of a $c$-function:
{\it i)} it has a maximum at the Gaussian fixed point, {\it ii)} it has
a saddle at the Wilson fixed point, and {\it iii)} there is only one
flow connecting both points ( we have not normalized the $c$-funtion to
one for the Gaussian fixed point as often done in the literature).
Naturally, this description corresponds to our particular
parametrization in terms of $m$ and $\lambda$, which implicitly carries
a choice of subtraction point.  The variation of ${\cal C}$ between
fixed points is reparametrization invariant and its positivity amounts
to physical irreversibility of the flow.  A contour plot of $\cal C$ for
$d=3$ is given in Fig. 3, which depicts the space of theories in the
basis given by $m$ and $\lambda$ as a hilly landscape.

For the sake of completeness, let us comment that the first mention of
irreversibility of the renormalization group flow was spelled out in the
context of perturbation theory by Wallace and Zhia \wallace.  Later,
Zamolodchikov\zamolodchikov\ proved a theorem in two dimensions, the
$c$-theorem, which relates the irreversibility of the flows to the basic
assumption of unitarity in the Hilbert space of the theory.  Several
authors\osborn\ have subsequently come to the conclusion that a similar
theorem holds in any dimension in perturbation theory.  More generally,
any expansion where the space of theories is reduced to a manifold in a
space of couplings will accomodate a $c$-theorem.  Our setting in this
Letter does not clearly fall into this category, due to the appearance
of rational functions of the couplings in Eq.  (5.4), and the explicit
construction of the $c$-function, though to first non-trivial order,
might be of relevance.

A systematic approach to the irreversibility of the renormalization
group flow in the projected Wegner-Houghton equation should rely upon a
computation of Zamolodchikov's metric ({\it i.e.} all two-point
correlators between composite operators in the theory).  This will
require an exact renormalization group equation for the generating
functional equipped with a source for composite scalar fields.

\bigskip

\noindent{\bf Acknowledgments}

We thank A. Cappelli for useful comments and discussions.\bigskip
\eject

\listrefs

\bigskip\bigskip

\noindent{\bf Figure Captions}

\bigskip {\it Fig}. 1. Critical exponents for $2.9 \le d \le 4$
corresponding to the relevant, marginal and the first two irrelevant
operators in the $M=7$ approximation. \bigskip

{\it Fig}. 2. $d=3$ Renormalization group flows projected on mass and
quartic coupling subspace in the $M=3$ approximation. $c_1$ is plotted
on the $x$-axis and $c_3$ on the $y$-axis. \bigskip

{\it Fig}. 3. $c$-function contour of Eq.  (5.6).  The Gaussian point is
at the top of the hill $(0,0)$, whereas the Wilson point lies on the
saddle $(-1/7,\sqrt{24}/7)$. \vfill
\end